\documentclass[twocolumn,showpacs,preprintnumbers,superscriptaddress,amsmath,amssymb,PRC]{revtex4}
\usepackage{graphicx}
\usepackage{dcolumn}
\usepackage{float}
\usepackage{mathptmx, courier, pifont}
\usepackage[scaled=0.92]{helvet}
\usepackage[T1]{fontenc}
\usepackage{textcomp}
\usepackage{color}

\begin{document}
\title{Identification of the Lowest $T=2$, $J^{\pi=}0^+$ Isobaric Analog State in $^{52}$Co and Its Impact on the Understanding of $\beta$-Decay Properties of $^{52}$Ni}

\author{X.~Xu}
\affiliation{Key Laboratory of High Precision Nuclear Spectroscopy and Center for Nuclear Matter Science, Institute of Modern Physics, Chinese Academy of Sciences, Lanzhou 730000, People's Republic of China}
\author{P.~Zhang}
\affiliation{Key Laboratory of High Precision Nuclear Spectroscopy and Center for Nuclear Matter Science, Institute of Modern Physics, Chinese Academy of Sciences, Lanzhou 730000, People's Republic of China}
\affiliation{Graduate University of Chinese Academy of Sciences,
	Beijing, 100049, People's Republic of China}
\author{P.~Shuai}
\affiliation{Key Laboratory of High Precision Nuclear Spectroscopy and Center for Nuclear Matter Science, Institute of Modern Physics, Chinese Academy of Sciences, Lanzhou 730000, People's Republic of China}
\author{R.~J.~Chen}
\affiliation{Key Laboratory of High Precision Nuclear Spectroscopy and Center for Nuclear Matter Science, Institute of Modern Physics, Chinese Academy of Sciences, Lanzhou 730000, People's Republic of China}
\author{X.~L.~Yan}
\affiliation{Key Laboratory of High Precision Nuclear Spectroscopy and Center for Nuclear Matter Science, Institute of Modern Physics, Chinese Academy of Sciences, Lanzhou 730000, People's Republic of China}
\author{Y.~H.~Zhang}\thanks{Corresponding author. Email address: yhzhang@impcas.ac.cn}
\affiliation{Key Laboratory of High Precision Nuclear Spectroscopy and Center for Nuclear Matter Science, Institute of Modern Physics, Chinese Academy of Sciences, Lanzhou 730000, People's Republic of China}
\author{M.~Wang}\thanks{Corresponding author. Email address: wangm@impcas.ac.cn}
\affiliation{Key Laboratory of High Precision Nuclear Spectroscopy and Center for Nuclear Matter Science, Institute of Modern Physics, Chinese Academy of Sciences, Lanzhou 730000, People's Republic of China}
\author{Yu.~A. Litvinov}\thanks{Corresponding author. Email address: y.litvinov@gsi.de}
\affiliation{Key Laboratory of High Precision Nuclear Spectroscopy and Center for Nuclear Matter Science, Institute of Modern Physics, Chinese Academy of Sciences, Lanzhou 730000, People's Republic of China}
\affiliation{GSI Helmholtzzentrum f\"{u}r Schwerionenforschung,
	Planckstra{\ss}e 1, 64291 Darmstadt, Germany}
\author{H.~S.~Xu}
\affiliation{Key Laboratory of High Precision Nuclear Spectroscopy and Center for Nuclear Matter Science, Institute of Modern Physics, Chinese Academy of Sciences, Lanzhou 730000, People's Republic of China}
\author{T.~Bao}
\affiliation{Key Laboratory of High Precision Nuclear Spectroscopy and Center for Nuclear Matter Science, Institute of Modern Physics, Chinese Academy of Sciences, Lanzhou 730000, People's Republic of China}
\author{X.~C.~Chen}
\affiliation{Key Laboratory of High Precision Nuclear Spectroscopy and Center for Nuclear Matter Science, Institute of Modern Physics, Chinese Academy of Sciences, Lanzhou 730000, People's Republic of China}
\affiliation{GSI Helmholtzzentrum f\"{u}r Schwerionenforschung,
	Planckstra{\ss}e 1, 64291 Darmstadt, Germany}
\author{H.~Chen}
\affiliation{Key Laboratory of High Precision Nuclear Spectroscopy and Center for Nuclear Matter Science, Institute of Modern Physics, Chinese Academy of Sciences, Lanzhou 730000, People's Republic of China}
\affiliation{Graduate University of Chinese Academy of Sciences,
	Beijing, 100049, People's Republic of China}
\author{C.~Y.~Fu}
\affiliation{Key Laboratory of High Precision Nuclear Spectroscopy and Center for Nuclear Matter Science, Institute of Modern Physics, Chinese Academy of Sciences, Lanzhou 730000, People's Republic of China}
\affiliation{Graduate University of Chinese Academy of Sciences,
	Beijing, 100049, People's Republic of China}
\author{S.~Kubono}
\affiliation{Key Laboratory of High Precision Nuclear Spectroscopy and Center for Nuclear Matter Science, Institute of Modern Physics, Chinese Academy of Sciences, Lanzhou 730000, People's Republic of China}
\author{Y.~H.~Lam}
\affiliation{Key Laboratory of High Precision Nuclear Spectroscopy and Center for Nuclear Matter Science, Institute of Modern Physics, Chinese Academy of Sciences, Lanzhou 730000, People's Republic of China}
\author{D.~W.~Liu}
\affiliation{Key Laboratory of High Precision Nuclear Spectroscopy and Center for Nuclear Matter Science, Institute of Modern Physics, Chinese Academy of Sciences, Lanzhou 730000, People's Republic of China}
\affiliation{Graduate University of Chinese Academy of Sciences,
	Beijing, 100049, People's Republic of China}
\author{R.~S.~Mao}
\affiliation{Key Laboratory of High Precision Nuclear Spectroscopy and Center for Nuclear Matter Science, Institute of Modern Physics, Chinese Academy of Sciences, Lanzhou 730000, People's Republic of China}
\author{X.~W.~Ma}
\affiliation{Key Laboratory of High Precision Nuclear Spectroscopy and Center for Nuclear Matter Science, Institute of Modern Physics, Chinese Academy of Sciences, Lanzhou 730000, People's Republic of China}
\author{M.~Z.~Sun}
\affiliation{Key Laboratory of High Precision Nuclear Spectroscopy and Center for Nuclear Matter Science, Institute of Modern Physics, Chinese Academy of Sciences, Lanzhou 730000, People's Republic of China}
\affiliation{Graduate University of Chinese Academy of Sciences,
	Beijing, 100049, People's Republic of China}
\author{X.~L.~Tu}
\affiliation{Key Laboratory of High Precision Nuclear Spectroscopy and Center for Nuclear Matter Science, Institute of Modern Physics, Chinese Academy of Sciences, Lanzhou 730000, People's Republic of China}
\affiliation{Max-Planck-Institut f\"{u}r Kernphysik, Saupfercheckweg 1, 69117 Heidelberg, Germany}
\author{Y.~M.~Xing}
\affiliation{Key Laboratory of High Precision Nuclear Spectroscopy and Center for Nuclear Matter Science, Institute of Modern Physics, Chinese Academy of Sciences, Lanzhou 730000, People's Republic of China}
\affiliation{Graduate University of Chinese Academy of Sciences,
	Beijing, 100049, People's Republic of China}
\author{J.~C.~Yang}
\affiliation{Key Laboratory of High Precision Nuclear Spectroscopy and Center for Nuclear Matter Science, Institute of Modern Physics, Chinese Academy of Sciences, Lanzhou 730000, People's Republic of China}
\author{Y.~J.~Yuan}
\affiliation{Key Laboratory of High Precision Nuclear Spectroscopy and Center for Nuclear Matter Science, Institute of Modern Physics, Chinese Academy of Sciences, Lanzhou 730000, People's Republic of China}
\author{Q.~Zeng}
\affiliation{Research Center for Hadron Physics, National Laboratory of Heavy Ion Accelerator Facility in Lanzhou and University of Science and Technology of China, Hefei 230026, People's Republic of China}
\affiliation{Key Laboratory of High Precision Nuclear Spectroscopy and Center for Nuclear Matter Science, Institute of Modern Physics, Chinese Academy of Sciences, Lanzhou 730000, People's Republic of China}
\author{X.~Zhou}
\affiliation{Key Laboratory of High Precision Nuclear Spectroscopy and Center for Nuclear Matter Science, Institute of Modern Physics, Chinese Academy of Sciences, Lanzhou 730000, People's Republic of China}
\affiliation{Graduate University of Chinese Academy of Sciences,
	Beijing, 100049, People's Republic of China}
\author{X.~H.~Zhou}
\affiliation{Key Laboratory of High Precision Nuclear Spectroscopy and Center for Nuclear Matter Science, Institute of Modern Physics, Chinese Academy of Sciences, Lanzhou 730000, People's Republic of China}
\author{W.~L.~Zhan}
\affiliation{Key Laboratory of High Precision Nuclear Spectroscopy and Center for Nuclear Matter Science, Institute of Modern Physics, Chinese Academy of Sciences, Lanzhou 730000, People's Republic of China}
\author{S.~Litvinov}
\affiliation{GSI Helmholtzzentrum f\"{u}r Schwerionenforschung,
	Planckstra{\ss}e 1, 64291 Darmstadt, Germany}
\author{K.~Blaum}
\affiliation{Max-Planck-Institut f\"{u}r Kernphysik, Saupfercheckweg 1, 69117 Heidelberg, Germany}
\author{G.~Audi}
\affiliation{CSNSM, Univ Paris-Sud, CNRS/IN2P3, Universit\'{e} Paris-Saclay, 91405 Orsay, France}
\author{T.~Uesaka}
\affiliation{RIKEN Nishina Center, RIKEN, Saitama 351-0198, Japan}
\author{Y. Yamaguchi}
\affiliation{RIKEN Nishina Center, RIKEN, Saitama 351-0198, Japan}
\author{T. Yamaguchi}
\affiliation{Department of Physics, Saitama University, Saitama 338-8570, Japan}
\author{A.~Ozawa}
\affiliation{Insititute of Physics, University of Tsukuba, Ibaraki 305-8571, Japan}
\author{B.~H.~Sun}
\affiliation{School of Physics and Nuclear Energy Engineering,
	Beihang University, Beijing 100191, People's Republic of China}
\author{Y.~Sun}
\affiliation{Department of Physics and Astronomy, Shanghai Jiao Tong University,
	Shanghai 200240, People's Republic of China} 
\author{A.~C.~Dai}
\affiliation{State Key Laboratory of Nuclear Physics and Technology, School of Physics, Peking University, Beijing 100871, People's Republic of China} 
\author{F.~R.~Xu}
\affiliation{State Key Laboratory of Nuclear Physics and Technology, School of Physics, Peking University, Beijing 100871, People's Republic of China} 

\begin{abstract}
	
Masses of $^{52g,52m}$Co were measured for the first time with an accuracy of $\sim 10$ keV, an unprecedented precision  reached for short-lived nuclei in the isochronous mass spectrometry. Combining our results with the previous $\beta$-$\gamma$ measurements of $^{52}$Ni, the $T=2$, $J^{\pi}=0^+$ isobaric analog state (IAS) in $^{52}$Co was newly assigned, questioning the conventional identification of IASs from the $\beta$-delayed proton emissions. Using our energy of the IAS in $^{52}$Co, the masses of the $T=2$ multiplet fit well into the Isobaric Multiplet Mass Equation. We find that the IAS in $^{52}$Co decays predominantly via $\gamma$ transitions while the proton emission is negligibly small. According to our large-scale shell model calculations, this phenomenon has been interpreted to be due to very low isospin mixing in the IAS.

\end{abstract}

\pacs{21.10.Dr, 27.40.+z, 29.20.db}

\maketitle


The concept of isospin was introduced by Heisenberg~\cite{Heis32} and developed by Wigner~\cite{Wigner37} to describe the charge independence of nuclear forces. This concept is being widely used in particle and nuclear physics~\cite{isospin1,isospin2}. 
Within the isospin formalism, a nucleus composed of $Z$ protons and $N$ neutrons 
has a fixed isospin projection of $T_z=(N-Z)/2$, while  
all states in the nucleus can have different total isospins $T \ge |T_z|$.
In other words, states of a given $T$ can occur in a set of isobaric nuclei with $T_z=T,T-1,...,-T$.
These states with the same $T$ and $J^{\pi}$ are called the isobaric analog states (IAS).
The states with $T=|T_z|$ are the ground states of the corresponding nuclei and the ones  
with $T>|T_z|$ are excited states, except for some odd-odd $N=Z$ nuclei~\cite{Janecke2002,AME2012}.
A set of IASs with fixed $A$ and $T$ are believed to have very similar structure 
and properties and to be energetically degenerated in the framework of isospin symmetry. 
This energy degeneracy is mainly altered due to the Coulomb interaction, the proton-neutron mass difference, and the charge-dependent forces of nuclear origin~\cite{bently2007}. 
In an isobaric multiplet, the masses of the IASs of a given $T$ can be described in first order approximation
by the famous quadratic Isobaric Multiplet Mass Equation IMME~\cite{Wigner57,Wein59,Bene79}
\begin{equation}
\ ME(A,T,T_z)=a(A,T)+b(A,T)T_z+c(A,T)T^2_z, \label{eq3}
\end{equation} 
where $a$, $b$, and $c$ are coefficients.
\par
Identification of IASs and determination of their basic properties, like energies, lifetimes and decay branching ratios, have long been an important research subject.
The latter is due to several motivations:
(i) extracted Coulomb displacement energies between neighboring IASs constrain nuclear structure theory
and allow for investigations of isospin-symmetry breaking effects of different origins 
(see Refs.~\cite{Nolen69,Shlomo78,Ben07} for  reviews); 
(ii) a complete set of masses for any $T\ge 1 $ isobaric multiplet in the $sd$- or $fp$-shell can be used to test the validity of the IMME~\cite{Blaum03,20Mg,Ni53a} 
as well as to extract information on the vector and tensor components of the isospin non-concerving forces~\cite{Ormand1989,Lam13};
(iii) precise mass values of the $T=1$ IASs are used, 
in combination with the associated super-allowed $0^+\rightarrow 0^+$ $\beta$ decay properties, 
to test the Conserved Vector Current hypothesis of the electroweak interaction~\cite{HT05,HT15}, 
which is an active research field for more than 50-years;
(iv) the analysis of the IASs provides accurate mass predictions for 
neutron-deficient nuclei yet inaccessible in experiments, which in turn are valuable, e.g., 
for modelling the astrophysical rp-process of nucleosynthesis~\cite{Sch06,Pari09}.       
\par
A compilation of data on the IASs throughout the nuclear chart can be found in Ref.~\cite{Ant97}. 
The $T=2$, $J^{\pi}=0^+$ IAS in $^{52}$Co was proposed in Refs.~\cite{Faux94,Dossat,Orrigo16} 
based on the data from $\beta$-delayed proton decay ($\beta$-$p$) of the $T_z=-2$ nucleus $^{52}$Ni. 
However its energy was excluded from the recent evaluation of the IASs since it significantly 
deviates from the value calculated with the IMME~\cite{Mac14}. 
\par
In this Letter, we report on the first measurement of the masses of ground state $^{52}$Co and its low-lying $(2^+)$ isomer.
Combined with data on $\beta$-delayed $\gamma$-decay ($\beta$-$\gamma$) of $^{52}$Ni~\cite{Dossat,Orrigo16},
this allowed us to determine the energy of the $T=2$ IAS in $^{52}$Co.
We show that the IAS decays predominantly through $\gamma$ de-excitation and thus
question the conventional way of IAS assignment based on the relative intensity of proton groups~\cite{cerny1977}.
\par
The experiment was performed at the Heavy Ion Research Facility in Lanzhou (HIRFL) and Cooler Storage Ring (CSR) accelerator complex. 
The high-energy part of the facility consists of a main cooler-storage ring (CSRm), 
operating as a heavy-ion synchrotron, and an experimental ring CSRe coupled to CSRm
by an in-flight fragment separator RIBLL2~\cite{Xia02}. 
Details of the experiment and data analysis can be found in Ref.~\cite{Tu11}. 
Only a brief outline is given in the following.
\par
A 467.91~MeV/u $^{58}$Ni$^{19+}$ primary beam from the CSRm 
was focused onto a $\sim $15~mm thick beryllium target placed in front of the in-flight fragment separator RIBLL2. The reaction products from projectile fragmentation of $^{58}$Ni emerged from the target at relativistic energies and mostly as bare nuclei. The charge-state distributions can be estimated with a specialized CHARGE code~\cite{CHARGE}. For instance, the calculated fraction of fully-ionized atoms for Co is 99.92$\%$. 
The fragments were selected and analyzed~\cite{Geis92} by RIBLL2. 
A cocktail beam of $10 \sim 20 $ particles per spill were injected into the CSRe. 
The CSRe was tuned into the isochronous ion-optical mode ~\cite{Tu11,Haus00} with the transition point at $\gamma_t=1.4$. 
The primary beam energy was selected according to the LISE++ simulations~\cite{Tar08} such that the $^{52}$Co$^{27+}$ ions had the most probable velocity with $\gamma=\gamma_t$ at the exit of the target. 
Both RIBLL2 and CSRe were set to a fixed magnetic rigidity of $B\rho=5.8574$~Tm to allow for an optimal transmission of the $T_z=-1$ nuclides centered on $^{52}$Co. 
In order to increase the mass resolving power, a 60~mm wide slit was introduced in the dispersive straight section of the CSRe
to reduce the momentum spread of the secondary beams in the CSRe.
\par
\begin{figure} [t]
	\includegraphics[angle=0,width=8.5cm]{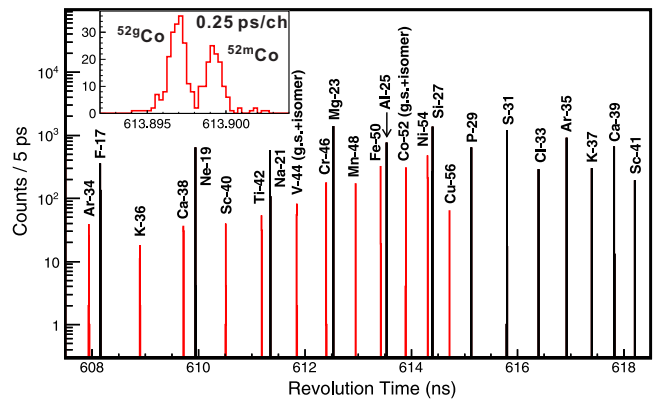}
	\caption{(Colour online) Part of the revolution time spectrum zoomed in at a time window of $608~{\rm ns} \le t \le 619$~ns. The red and black peaks represent the $T_z=-1$ and $-1/2$ nuclei, respectively. 
	The insert shows well-resolved peaks of the ground- and $(2^+)$ isomeric- states of $^{52}$Co.
		\label{Fig01}}
\end{figure}

The revolution times of the stored ions were measured using a timing detector~\cite{Mei10} installed inside the ring aperture. 
Each time an ion passed through the carbon foil of the detector, a timing signal was generated and recorded by a fast digital oscilloscope. 
By analyzing the timing signals the revolution time for each ion was obtained, and finally   
the revolution-time spectrum was created by accumulating all the events. 
Figure~\ref{Fig01} shows a part of the spectrum measured in this work and zoomed in at a time window of $608~{\rm ns} \le t \le 619$~ns. 
The identification of the peaks was done in the same way as in Ref.~\cite{Tu11}. 
The clearly resolved ground- ($^{\rm 52g}$Co) and low-lying isomeric- ($^{\rm 52m}$Co) states of $^{52}$Co are shown in the insert.

The analysis of data was conducted according to the procedure described in \cite{Ni53a,Tu11,Tu-PRL,Shuai-PLB}.
The measured revolution times of $^{52}$Co and its $(2^+)$ isomer were fitted using the unbinned maximum likelihood method. The mean revolution times of the ground and isomeric states of $^{52}$Co were determined to be 613.89685(5) ns and 613.89935(7) ns, respectively.
The corresponding mass values were then determined via the interpolation of the mass calibration function.
 
\par
The mass excesses, $ME=(m-A\cdot u)c^2$, 
are directly measured 
in this work to be $ME(^{\rm 52g}{\rm Co})=-34361(8)$~keV and $ME(^{\rm 52m}{\rm Co})=-33974(10)$~keV, respectively. 
These values are by 371(200)~keV and 364(220)~keV, respectively, lower 
than the extrapolated ones 
in the latest Atomic-Mass Evaluation (${\rm AME}{^\prime}$12)~\cite{AME2012}. 
The isomer excitation energy equals to $E_x=387(13)$~keV, 
which is very close to $E_x=378$~keV of the $2^+$ isomer in the mirror nucleus $^{52}$Mn~\cite{52Mn73}. 
\par
The $\beta$-$p$ and $\beta$-$\gamma$ decay of the $T_z=-2$ nucleus $^{52}$Ni 
was investigated in Refs.~\cite{Faux94,Dossat,Orrigo16},
where a strong proton peak with decay energy of $Q_p=1352$ keV and 
a $\gamma$ cascade of 2407- and 141-keV sequential transitions were observed.
In the following we use the most recent data from Ref.~\cite{Orrigo16}.
\par
As conventionally done, the strongest 1352-keV proton peak was first assigned in 
Ref.~\cite{Faux94} and then adopted in Refs.~\cite{Dossat,Orrigo16} 
as being due to the de-excitation of the expected IAS in $^{52}$Co 
to the ground state of $^{51}$Fe, thus giving the mass excess of the IAS of $ME(^{52}{\rm Co}^{\rm IAS})=-31561(14)$~keV.
\par
The coincident $2407(1)$-keV and $141(1)$-keV $\gamma$ rays 
de-excite the IAS feeding the $(2^+)$ isomer in $^{52}$Co (see Refs.~\cite{Dossat,Orrigo16}).
Since the masses of $^{\rm 52g,52m}$Co have been measured in this work, 
the $ME(^{52}{\rm Co}^{\rm IAS})$ could {\it independently} be determined to be $ME(^{52}{\rm Co}^{\rm IAS})=-31426(10)$~keV.
\par
The two $ME$ values disagree by 135(17)~keV.
This $\sim8\sigma$ deviation can not be due to
experimental uncertainties and calls for a different interpretation of available data,
namely that the observed 2407-keV $\gamma$ and 1352(10)-keV proton
in the decay of $^{52}$Ni~\cite{Dossat,Orrigo16} are from two different excited states in $^{52}$Co.
\par
We emphasize that the same experiment \cite{Orrigo16} reports $\beta$-$p$ and $\beta$-$\gamma$ data of the $^{48}$Fe decay.
By using the proton-decay energy $Q_p=1018(10)$~keV from Ref.~\cite{Orrigo16} and 
$ME$($^{47}$Cr)$=-34561(7)$~keV from Ref.~\cite{AME2012}, 
$ME(^{48}{\rm Mn}^{\rm IAS})=-26254(12)$~keV can be deduced.
The mass of $^{48}$Mn was also measured in our present experiment.
By taking our $ME(^{48}$Mn)$=-29299(7)$~keV 
and the corresponding $\gamma$-ray energies from Ref.~\cite{Orrigo16},
we get $ME(^{48}{\rm Mn}^{\rm IAS})=-26263(8)$~keV.
We see that two $ME(^{48}{\rm Mn}^{\rm IAS})$ values from two decay channels 
of the IAS in $^{48}{\rm Mn}$ are in excellent agreement.
This agreement supports our approach in the analysis of the $^{52}$Co data.
\par
The absolute intensity of 42(10)\% for the 2407-keV $\gamma$ transition measured in $^{52}$Co is
much stronger than the 13.7(2)\% 1352(10)-keV proton emission~\cite{Orrigo16}.
Hence, it is reasonable to assign the former as from the IAS in $^{52}$Co with $ME=-31426(10)$~keV, 
and the latter as from a $1^+$ state with $ME=-31561(14)$~keV, 
which could be the analog $1^+$ state to the one identified in the mirror nucleus $^{52}$Mn~\cite{52Mn73,fujita15}. 
\par
The assignment of the $ME(^{52}{\rm Co}^{\rm IAS})=-31426(10)$~keV can further be tested by the IMME, see Eq.~(\ref{eq3}). A deviation to the quadratic form of the IMME can be quantified by adding a cubic term $d\times T_z^3$. By using the $ME(^{52}{\rm Co}^{\rm IAS})=-31561(14)$~keV, 
Dossat {\it et al.} found that the $d$-coefficient deviates significantly from zero.
They attributed this deviation to a misidentification of one of the states assigned to this isobaric multiplet. Recently, the experimental IASs from $T=1/2$ to $T=3$ have been evaluated and 
the associated IMME coefficients were investigated in Ref.~\cite{Mac14}.
The assigned $ME(^{52}{\rm Co}^{\rm IAS})=-31561(14)$~keV in Refs.~\cite{Faux94,Dossat,Orrigo16} had to be excluded from the IMME fit because the $c$-coefficient dramatically deviates from a smooth trend. In contrary, our $ME(^{52}{\rm Co}^{\rm IAS})=-31426(10)$~keV combined with known
$T=2$ IASs in $^{52}$Fe, $^{52}$Mn, and $^{52}$Cr fits well into the quadratic form of the IMME with a normalized $\chi_n= 1.37$. The corresponding calculated $d$-coefficient, $d=5.8(4.2)$, is compatible with zero within 1.4$\sigma$.    
\par
\begin{figure}[t]
        \centering
        \rotatebox[]{0}{\includegraphics[width=8.5cm,angle=0]{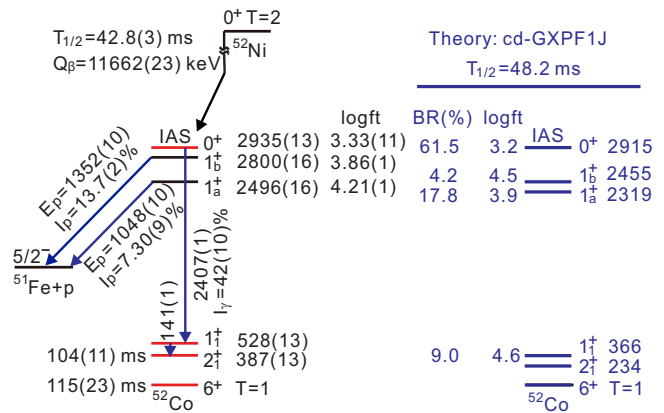}} 
        \caption{
		(Color online) Partial decay scheme of $^{52}$Ni (left) and theoretical level structure of $^{52}$Co (right). 
		Excitation energies are in keV. The theoretical branching ratios $(BR)$ and $log~ft$ values based on cd-GXPF1J are deduced from the \emph{present} $Q$ value. 
		The red levels are deduced from the ground-state mass of $^{52}$Co and the $\gamma$-ray energies from Ref.~\cite{Orrigo16}. 
		The black levels are determined from the $\beta$-$
		p$ data. }
		\label{Fig02}
       \end{figure}

   
Taking the newly assigned IAS in $^{52}$Co and 
the $\beta$-$p$ and $\beta$-$\gamma$ data of $^{52}$Ni~\cite{Dossat,Orrigo16}, 
we reconstructed the partial decay scheme of $^{52}$Ni as shown in Fig.~\ref{Fig02}. 
The $J^{\pi}$ assignments for the levels in $^{52}$Co are 
inferred from the analogous states in the mirror nucleus $^{52}$Mn~\cite{fujita15}. 
By using the IMME, the mass excess of $^{52}$Ni is predicted to be $ME(^{52}{\rm Ni})=-22699(22)$~keV and the $Q_{\rm EC}$ value of $^{52}$Ni is thus deduced to be $11662(23)$~keV.
The main modification in the present level scheme is that we attribute the 1352-keV proton 
to originate from the decay of the $1_b^+$ state rather than from the IAS. 
The excitation energies of the $1_a^+$ and $1_b^+$ states are calculated by subtracting the $ME(^{\rm 52g}{\rm Co})$ 
measured in this work from the $ME$ values deduced from the $\beta$-$p$ data. 
The $logft$ values are deduced~\cite{logft} for each individual $\beta$ transitions
according to this partial level scheme and the $\beta$-$p$ and $\beta$-$\gamma$ intensities given in Ref.~~\cite{Orrigo16}. 
The rather small $log~ft$ value of 3.33(11) to the IAS is consistent with the super-allowed Fermi decay of $^{52}$Ni. 
\par
As it is usually expected in the $sd$- and $fp$-shell neutron-deficient nuclei, 
the newly assigned IAS should proceed via a strong 1487(14)-keV proton emission to the ground state of $^{51}$Fe. However, such a proton peak was not observed in the high-statistics proton spectrum in Fig.~16 of Ref.~\cite{Orrigo16}. The strongest proton peak there is at 1352~keV and it does not show any visible broadening. Taking into account that the Double-Sided Silicon Strip Detector (DSSSD) used in Ref.~\cite{Orrigo16} had an energy resolution of 70~keV (FWHM), two nearby proton peaks with 135~keV energy difference would clearly be separated in the $\beta$-$p$ spectrum of $^{52}$Ni.
Hence, we conclude that the proton decay branch of the IAS in $^{52}$Co is negligibly small.
\par
This finding has important implications on the identification of the IASs 
in the study of $\beta$-delayed charged-particle emissions~\cite{Blank08}. 
It has been conventionally assumed that the IAS in a neutron-deficient nucleus decays mainly, when it is more than 1~MeV proton-unbound, via a proton emission due to a small isospin mixing~\cite{cerny1977,Blank08}.
Consequently, the strongest proton peak is often assigned as being from the IAS of a daughter nucleus of the $\beta$-$p$ precursor~\cite{Dossat,Blank08}. This identification may become unsafe in the $fp$-shell nuclei, e.g., in the $^{52}$Ni decay, if no other information is available. 
By inspecting Ref.~\cite{Dossat} we find that for several neutron-deficient $fp$-shell nuclei 
the $\beta$-$p$ strengths from the IASs are much weaker than the predictions of the super-allowed $\beta^+$ feeding. Therefore it is crucial to measure $\beta$-$\gamma$ data in order to make a firm identification of the IAS. Indeed, such a measurement has been performed recently on $^{53}$Ni~\cite{Su16} and the $T=3/2$ IAS in $^{53}$Co was found to be $\sim 70$ keV 
below the previously assigned IAS on the basis of $\beta$-$p$ emission data~\cite{Dossat}.       

\par
In this work we identified a new case, $^{52}$Co, in which the IAS decays 
by $\gamma$ transition rather than by proton emission although it is $\sim 1.5$~MeV proton-unbound. 
To understand this phenomenon and explore the details of the reconstructed partial decay scheme of $^{52}$Ni, we performed large-scale shell model calculations in the full $fp$-shell by using NuShellX@MSU code~\cite{NuShellX}. The isospin non-conserving (INC) Hamiltonian (hereinafter referred to as cd-GXPF1J) is constructed based on the isospin conserving Hamiltonian 
cd-GXPF1J~\cite{GXPF1J}, the Coulomb interaction, and the isovector single-particle energies (IVSPEs)~\cite{OrBr85} scaled as $\sqrt{\hbar \omega (A)}$~\cite{Lam13}. 
A quenching factor $q_F$=0.74 to the Gamow-Teller (GT) operator is employed to calculate the $\beta$-decay strength distribution.
The modern GXPF1J interaction has recently been used to reproduce various experimental 
Gamow-Teller (GT) strengths in the region close to $A=52$~\cite{fujita15}. 
The calculated results are shown in the right part of Fig.~\ref{Fig02} including the partial level structure of $^{52}$Co, $^{52}$Ni $\beta$-decay branching ratios, $BR$, and their $logft$ values.
Theoretical calculations agree well to the experiment. 
Furthermore, the theoretical half-life of 48.2~ms agrees well to the experimental value of 42.8(3)~ms \cite{AME2012}.

The proton- and $\gamma$-decay branches from the excited states of $^{52}$Co were calculated. 
The $\gamma$ widths, $\Gamma_\gamma$, 
were calculated by using the effective electromagnetic operators from Ref.~\cite{GXPF1A}.
The total proton width can be described as $\Gamma_p = \sum_{nlj}C^2S(nlj)\Gamma_{sp}(nlj)$, where $C^2S(nlj)$ is a single-particle spectroscopic factor, and $\Gamma_{sp}$ denotes a single-particle width for the proton emission from an $(nlj)$ quantum orbital~\cite{MacFrench}. The $\Gamma_{sp}$ is obtained from proton scattering cross sections described by the Woods-Saxon potential~\cite{YiHua2016,WSPOT}. Two important results are obtained: 
\par
(1) Our theoretical calculations predict a super-allowed $\beta$ transition of $^{52}$Ni to its 
IAS in $^{52}$Co with a branching ratio of $61.5\%$. 
Furthermore, three Gamow-Teller transitions are predicted to feed the $1^+$ states below IAS 
with branching ratios of $4\%$ through 18\% (see Fig.~\ref{Fig02}). 
The four calculated $\beta$-decay branches sum to 92.5\% of the total decay strength of $^{52}$Ni, 
which is in good agreement with the expected $\beta$-decay spectrum deduced 
from the $^{52}$Cr($^3$He,t)$^{52}$Mn charge exchange reaction~\cite{fujita15}.
Especially the relative strengths of the three Gamow-Teller transitions have been well reproduced.  
\par
(2) The calculations show that the total proton width of the IAS, $\Gamma_p^{\textnormal{IAS}}$, is 0.0001~eV, 
which is three orders of magnitude smaller than $\Gamma_\gamma^{\textnormal{IAS}}=0.25$~eV. 
This indicates that the IAS in $^{52}$Co decays predominantly via $\gamma$ transitions.
The proton emission should be orders of magnitude weaker than the $\gamma$ transitions 
and is thus unlikely to be observed experimentally. 
In fact, the $\beta$-$p$ emission from IAS is isospin forbidden, 
and the observation of such a proton emission is usually attributed to 
the isospin mixing of the IAS with the nearby $T=1, J^{\pi}=0^+$ states. 
In our shell model calculations, the closest $0^+$ state is predicted to be 168~keV below the IAS 
and the isospin mixing imposed from this $0^+$ to the IAS is calculated to be merely 0.23\%. 
The small isospin mixing indicates that no observable proton emission from the IAS is expected, 
which is consistent with our re-constructed decay scheme of $^{52}$Ni (see Fig.~\ref{Fig02}). 
Concerning the ${1^+_a}$ and ${1^+_b}$ states, the total proton widths are 0.6~eV and 37.8~eV, respectively, 
which are orders of magnitude larger than the $\gamma$ widths 
of $\Gamma_\gamma (1_a^+)=0.05$~eV and $\Gamma_\gamma (1_b^+)=0.04$~eV, respectively. 
This is again consistent with experiment that both $1^+$ states de-excite predominantly 
via proton emission and the $\gamma$ transitions were too weak to be observed~\cite{Dossat,Orrigo16}. 
\par
Although our shell model calculations provide an overall consistent interpretation of all available data on the $\beta$-decay of $^{52}$Ni, 
there are, however, three remaining open questions: 
\par
(1) the measured intensity of 7.3\% for the 1048-keV proton is much weaker than the predicted $\beta$ feeding of 17.8\%;
\par
(2) the intensity of 13.7\% for the 1352~keV proton emission is much higher than the predicted $\beta$ feeding of 4.2\%;
\par
(3) the intensity of 42\% for the 2407-keV $\gamma$ transition is lower than the predicted Fermi $\beta$ feeding of 61.5\%.  
\par
The first point may be caused by the un-observed $\gamma$ rays or protons de-exciting the $1^+_a$ level. The last two points may be interpreted, at least qualitatively, by assuming an ${\rm IAS}\rightarrow 1^+_b$ transition via $\gamma$ and internal electron conversion. 
Such an exotic $\beta$-delayed $\gamma$-$p$ decay has been observed in its neighboring nucleus $^{56}$Zn~\cite{56Zn}, although the comparable ${\rm IAS}\rightarrow 1^+_b$ 
decay branching in $^{52}$Co can not be predicted from our theoretical calculations. 
\par
The questions raised above vitalize us to propose an alternative scenario based on the hypothesis that there would exist a low-lying spherical state in $^{51}$Fe to which the IAS of $^{52}$Co may decay via proton emission. 
On the one hand, the ground-state of $^{52}$Ni is thought to be spherical due to its semi-magic character. Hence the IAS in $^{52}$Co should also be spherical according to the isospin symmetry. On the other hand, the rotation-like bands have been observed in $^{50,51,52}$Fe~\cite{50Fe01,51Fe00,52Fe98} indicating that the ground states of these isotopes are slightly deformed. Thus, the proton emission from the spherical $T=2$ IAS in $^{52}$Co to the 
$T=1/2$ deformed states in $^{51}$Fe are hindered not only by the isospin selection rules 
but also by the shape changes between the final and initial nuclear states, causing the proton emission from IAS to be less likely. However, if a shape coexistence in $^{51}$Fe is considered, 
it is possible that the soft nucleus $^{51}$Fe has in its Potential Energy Surface (PES) a second minimum with a nearly-spherical shape. To check this, we have performed PES calculations~\cite{XuPES} for $^{51}$Fe. 
Apart from the deformed minimum at $(\beta_2,\gamma)=(0.16,-10^{\circ})$ corresponding to the $5/2^-$ ground state, there exists a shallow -- nearly-spherical -- minimum $\sim 200$~keV above the deformed one. If such a spherical minimum exists in $^{51}$Fe, one may attribute the 1352-keV protons, or part of them, to be originated from the decay of the IAS to the states in the second minimum of $^{51}$Fe. Consequently, the intensity imbalances raised in the questions above could -- at least partly -- be solved.   

In conclusion, $^{58}$Ni projectile fragments were addressed by the isochronous mass spectrometry at HIRFL-CSR. Precision mass excess values for $^{52g}$Co and its low-lying (2$^+$) isomer have been precisely measured for the first time. Combining our new results with the literature $\beta$-$\gamma$ measurements of $^{52}$Ni, 
the energy of the $T=2$ isobaric analog state in $^{52}$Co was determined to be 135 keV higher 
than previously assumed on the basis of the $\beta$-$p$ data from the $^{52}$Ni decay studies. 
With this new IAS assignment, the mass excesses of the four members of the $A=52$, $T=2$ isobaric multiplet are found to be consistent with the quadratic form of the IMME. 
Furthermore, a remarkably different decay scheme of $^{52}$Ni could be constructed,
in which the proton group with the highest relative intensity~\cite{Dossat,Orrigo16} 
corresponds to the decay from the 1$^+$ excited state in $^{52}$Co and {\it not} from the 0$^+$, $T=2$ IAS state. This finding has important implications on the identification of the IASs from $\beta$-delayed charged-particle emission studies.
The newly determined level scheme of $^{52}$Co is consistent with its mirror nucleus $^{52}$Mn 
and can well be reproduced by large-scale shell model calculations using an isospin non-conserving Hamiltonian. Our theoretical calculations indicate that the isospin mixing in the 0$^+$, $T=2$ state in $^{52}$Co is extremely low, thus leading to a negligibly small proton emission from this state. An alternative scenario based on a possible shape coexistence is proposed 
to account for the remaining intensity imbalances observed between experiment and shell model calculations. Further experiments aiming at comprehensive studies of this interesting phenomenon are required. 
\par
The authors would like to thank Nicolas Winckler for his invaluable help in advanced data analysis methods and Nadezda A. Smirnova for stimulating discussions on the theoretical details. We thank also the staffs in the accelerator division of IMP for providing stable beam. 
This work is supported in part by the Major State Basic Research Development Program of China (Contract No. 2013CB834401), the NSFC grants U1232208, U1432125, 11205205, 11035007, 11235001, 11320101004, 11575007, the Chinese Academy of Sciences, and the Helmholtz-CAS Joint Research Group (HCJRG-108). K.B. acknowledge support by the Nuclear Astrophysics Virtual Institute (NAVI) of the Helmholtz Association. Y.H.L. thanks the support from the Ministry of Science and Technology of China (Talent Young Scientist Program) and from the China Postdoctoral Science Foundation (2014M562481). X.X. thanks the support from CAS "Light of West China" Program. 
\par

\end{document}